\documentclass[preprint,showpacs]{revtex4}
\usepackage{amsmath}
\usepackage{graphicx}
\usepackage{epstopdf}
\usepackage{newlfont}

\begin{document}
\title{Dynamical Structure Factors in Quantum Many-Body Systems from Quantum Monte Carlo Calculations}
\author{A. Roggero}
\affiliation{Dipartimento di Fisica, Universit\`a di Trento, via Sommarive, 14 I-38123 Trento, Italy}
\affiliation{ INFN - Istituto Nazionale di Fisica Nucleare, Gruppo Collegato di Trento,
  I-38123 Trento, Italy}
\email{roggero@science.unitn.it}
\author{F. Pederiva}
\affiliation{Dipartimento di Fisica, Universit\`a di Trento, via Sommarive, 14 I-38123 Trento, Italy}
\affiliation{ INFN - Istituto Nazionale di Fisica Nucleare, Gruppo Collegato di Trento,
  I-38123 Trento, Italy}
\email{pederiva@science.unitn.it}
\author{G. Orlandini} 
\affiliation{Dipartimento di Fisica, Universit\`a di Trento, via Sommarive, 14 I-38123 Trento, Italy}
\affiliation{ INFN - Istituto Nazionale di Fisica Nucleare, Gruppo Collegato di Trento,
  I-38123 Trento, Italy}
\email{orlandin@science.unitn.it}

\begin{abstract}
An {\it ab-initio} method for determining the dynamical structure 
function of an interacting many--body quantum system 
has been devised by combining a generalized integral transform method with Quantum Monte Carlo methods. 
As a first application, the coherent and, separately, the incoherent excitation spectrum of bulk atomic $^{4}$He has been computed, 
both in the low and intermediate momentum range. The peculiar form of the kernel in the integral transform of 
the dynamical structure function allows 
to predict, without using any model, both position and width 
of the collective excitations in the maxon--roton region, as well as  the second collective peak. A prediction of the dispersion of the single--particle modes described by the incoherent part is also presented. 
\end{abstract}
\pacs{02.70.Ss,67.25.dt,67.25.D-}
\maketitle

Many important physical properties of matter, from viscosity to magnetic susceptibility, are closely 
related to the underlying microscopic dynamics. The computation of such quantities is routinely performed for 
classical systems. Moreover, thanks to the growing computational capabilities,
we currently have a number of methods capable of a true $ab$-$initio$ treatment of the ground state of 
quantum many-body systems. However, the detailed study of dynamical properties has been yet elusive. 

In this letter we will focus on the problem of extending  in a consistent and reliable way 
a class of Quantum Monte Carlo (QMC) algorithms to make it possible to determine the Dynamical Structure Function (DSF) 
for a generic quantum many-body system and a generic excitation operator. 
Such an extension is based on the use of the Integral Transform (IT) technique with generalized kernels. 
The amount and quality of information that can be extracted by the proposed scheme is illustrated 
in the application to the study of the coherent and incoherent density excitation spectrum in $^{4}$He.

At present, QMC calculations provide benchmark results for the study of a huge 
variety of many-body systems ranging from quantum chemistry, physics of ultra-cold gases, 
and nuclear physics. 
The most dramatic limitation  of QMC methods is their inability to treat 
dynamical properties in a symilarly reliable way.
This failure is essentially due to the fact that QMC works in imaginary-time rather than in real time.  
This implies that quantities that do not directly translate into imaginary-time language, when analytically continued 
to real time, are affected 
by statistical noise that can be hardly reduced, making calculations unfeasible.

The mainstream approach to the problem is to attempt a numerical inversion starting from the 
imaginary-time autocorrelation function, i.e. the Laplace transform of the DSF.
As is well known, such an inversion is an exponentially ill-posed problem~\cite{Ha52} 
and sophisticated regularization techniques~\cite{TiA77,Ta79} are needed to correctly extract the physical information.
Due to the infinite range of the kernel function, the extraction of such informations is 
extremely delicate, and many details of the structure of the excitation spectrum can be missed.
As an example, one of the most powerful inversion schemes, the Maximum 
Entropy Method~\cite{Ja63,SiS90}, cannot resolve the (measured) double peaked structure of 
$S(\mathbf{q},\omega)$  in superfluid $^4$He~\cite{Cep-PIMC}, corresponding to a higher energy collective roton mode. In a recent paper this structure was eventually resolved inverting the imaginary time correlation function in a Path Integral Ground State calculation by using a falsification method based on  a genetic algorithm \cite{Vitali10}.

For strongly interacting  few-body systems the problem of computing various DSF
of density and current excitations is solved by using a generalized integral transform approach,
i.e. the Lorentz Integral 
Transform (LIT) method~\cite{EfL94,EfL07}.
The success of this approach, applied in nuclear physics, lays in the specific choice of 
the Lorentzian function as kernel of the IT. 
On the one hand this choice allows to calculate the transform with bound state techniques, 
even when the response is defined in the continuum, however avoiding its discretization.
On the other hand, and  most important, the fact that the kernel is a  representation of the $\delta$-function 
allows for a reliable and stable inversion.
So far, the application of this technique has been limited to a small number of particles (up to N=6~\cite{BaM02}
and 7~\cite{BaA04}), due to the huge computational costs of the diagonalizations needed to calculate 
the LIT. In the following we will discuss how to extend the idea of the LIT to many-body systems,
by developing a QMC equivalent.


At zero temperature the contribution to the response of a system of interacting particles due 
to a perturbative probe transferring momentum $\mathbf q$ and energy $\omega$ to it, can be expressed using a spectral representation
\begin{eqnarray}
\label{eq:res2}
S_{\hat O}(\mathbf q,\omega)&=\sum_{\nu} 
|\langle \Psi_{\nu} \lvert \hat{O}( \mathbf q) \rvert \Psi_{0} \rangle|^2 \delta (E_{\nu}-\omega ) \\
& = \langle \Psi_0 \lvert \hat{O^{\dagger}}( \mathbf q) \delta ( \hat{H} - \omega ) \hat{O} ( \mathbf q)\rvert \Psi_0 \rangle
\end{eqnarray}
where $\rvert \Psi_{0} \rangle$ is the ground state of the system, $\rvert \Psi_{\nu} \rangle$ are 
the final states of the reaction, $\hat{O}$ is an excitation operator, $\delta ( \hat{H} - \omega )$ 
is the spectral-density of the hamiltonian and the summation is extended to all discrete 
and continuum spectrum states in the set.

The cost of a direct calculation of $S_{\hat O}(\mathbf q,\omega)$ becomes rapidly prohibitive as the number 
of particles or the energy transfer $\omega$ increases. The latter problem is due to the fact 
that to account for continuum states one would need to solve the many-body scattering problem. 

One can instead consider an integral transform of $S_{\hat O}(\mathbf q,\omega)$  
with a generic kernel $K(\sigma,\omega)$ 
\begin{equation}
\label{eq:tr}
\Phi(\mathbf q,\sigma) = \int K(\sigma,\omega)\, S_{\hat O}(\mathbf q,\omega)\,d\omega.
\end{equation}
The substitution of the expression \eqref{eq:res2} for $S_{\hat O}(\mathbf q,\omega)$ yields:
\begin{eqnarray}
\label{eq:sm}
\Phi(\mathbf q,\sigma)&=& \sum_{\nu}  \langle\Psi_0 \lvert  \hat{O^{\dagger}}(\mathbf{q})\lvert \Psi_{\nu}
\rangle K(\sigma,\omega) \langle \Psi_{\nu} \lvert \hat{O}(\mathbf{q}) \rvert \Psi_0 \rangle \nonumber\\
&=& \langle \Psi_0 \lvert \hat{O^{\dagger}}(\mathbf{q}) \,K(\sigma,\hat{H}) \, \hat{O}(\mathbf{q}) 
\rvert \Psi_0 \rangle  
\end{eqnarray}
Equation \eqref{eq:sm} can be viewed as a {\it generalized sum rule} which depends on a continuous parameter $\sigma$.
Provided that the kernel and the excitation operator have suitable analytic properties, 
the right hand side of Eq.~\eqref{eq:sm} can be calculated using bound-state type methods. This is the case both 
for the Stieltjes kernel~\cite{Ef85,EfL93}, and 
for the Lorentz kernel. However, while in the former case the inversion of the transform is as problematic 
as in the case of the Laplace kernel, in the latter case even a rather simple  regularization procedure 
allows to obtain accurate and  stable results.
The reason can be easily understood. In the case of the Laplace or the Stieltjes kernels 
the information about 
$S_{\hat O}(\mathbf q,\omega) $ in the $\omega$ domain is spread in a large $\sigma$ domain. On the contrary 
the Lorentz kernel, as well as any function that is a $\delta$-function representation, 
keeps that information in an arbitrarily narrow $\sigma$ domain, governed  by the width of the kernel. 
(In the $\delta$ function limit of the kernel  no inversion would be needed!)


The number of $\delta$-function representations is large. However, very few have a practical implementation.
In the past the use of Gaussian kernels has been investigated in different fields from condensed matter 
(\cite{Munehisa},\cite{Korn}) to non perturbative QCD~\cite{Bertelman,OrlandiniQCD} with limited results.
Here the idea is 
to recast one possible $\delta$-function representation in the imaginary-time propagation language, typical of 
QMC methods, in a similar way as was suggested in Ref.~\cite{OrL10}. 

Consider the following family of integral kernels built out of the so-called Sumudu transform:
\begin{equation}
\label{eq:NEW4}
K(\sigma,\omega)=\frac{N}{\sigma}\left[ e^{-\mu\frac{\omega}{\sigma}}-e^{-\nu\frac{\omega}{\sigma}} \right]^P,
\end{equation}
where 
\begin{equation}
\mu=\frac{ln[b]-ln[a]}{b-a}a\,;\,\,\,\,\,\,\,
\nu=\frac{ln[b]-ln[a]}{b-a}b\,,
\end{equation}
and the parameters $P,a,b$ are integer numbers  with $b>a$. The normalization constant $N$ is a function of $P,a,b$
such that $\int d\omega K(\sigma,\omega)=1$.

The efficiency of this transform is displayed by the fact that the kernel function  converges to a delta function
$\delta(\omega-\sigma)$ in the $P \to \infty$ limit, independent on the choice of $a$ and $b$. 
For a finite $P$ the kernel $\sigma K_P(\sigma,\omega)$ is still centered around $\omega=\sigma$ but has a 
finite width that depends on  $P$. These properties make the choice of the resolution and the energy range of interest
extremely flexible, similarly to the Lorentian kernel.

Using a binomial expansion, and rewriting powers as exponential functions, leads to a more 
transparent form of the kernel:
\begin{equation}
\begin{split}
K_P(\sigma,\omega)& =\frac{N}{\sigma} \sum^{P}_{k=0} {
\begin{pmatrix}
P\\
k
\end{pmatrix}
} (-1)^k e^{-\ln(b/a)[\frac{a}{b-a}P+k]\frac{\omega}{\sigma}}. 
\end{split}
\end{equation}
By operating the substitution $\omega \to \hat{H}$ according to Eq. (\ref{eq:sm}), we are lead to a simple 
linear combination of imaginary-time propagators ($\hbar$=1), taken at imaginary-time points 
$\tau_{Pk}=\ln(b/a)[\frac{a}{b-a}P+k]
/\sigma$.
Projection QMC methods are all based on the implementation of such an 
imaginary-time propagation. The underlying idea is to solve the corresponding integral equation:
\begin{equation}
\Psi(R,\tau)=\int\; dR' G(R,R',\tau)\Psi(R',0).
\end{equation}
This can be achieved, for instance, sampling a representation in coordinate space of the Green's 
function $G(R,R',\tau)$ to propagate a set of configurations representing in turn an expansion of 
the function $\Psi(R,\tau)$ in eigenstates of the position operator (Diffusion Monte Carlo methods). 
Alternatively, it is possible to break up the Green's Function in a product of short time propagators 
in coordinate space:
\begin{equation}
\begin{array}{rcl}
\displaystyle
\Psi(R,\tau)&=&\int dR'\dots dR^{n} G(R,R^{n},\Delta \tau)\times  \\
&&\times G(R^{n},R^{n-1},\Delta \tau) \cdots G(R'',R',\Delta \tau)\psi(R',0),
\end{array}
\end{equation}
with $\tau = n\Delta \tau$. This formulation is implemented in the so-called 
Path-Integral Ground State methods~\cite{PIGS}, and in the Reptation Monte Carlo (RMC) 
algorithm~\cite{Moroni98}, where the whole path $\{ R,R',R'',\dots,R^{n}\}$ is sampled 
from the product of the short-time propagators $G$, possibly modified with the use of 
a suitable importance function $\Psi_{T}$ to be determined in a variational calculation.
The fact that estimating $\Phi(\sigma)$ reduces to the computation of an imaginary time 
correlation function makes this second formulation more convenient and straightforward. 
In particular, path-based methods yield estimates that never depend on the (necessary) 
importance function used to improve the convergence of the calculation.

In order to evaluate the transform \eqref{eq:tr} within a QMC approach, we need to compute the 
imaginary-time correlation function, and then construct the corresponding linear combinations.
A small width of the kernel might be achieved either using a large value of $P$ or by reducing the value of the ratio $b/a$. In both cases this would require the 
computation of the imaginary time correlation function for long imaginary time.
This might indeed be a serious complication. However, a kernel
that has a width smaller or comparable to the distance between the typical structures of the DSF
is in principle sufficient to extract useful informations from the transform. We have also confirmed 
this in  numerical tests. In the application described below we used the kernel $\sigma K_P(\sigma,\omega)$ with the typical values of $P=2$, $a=1$, and $b=2$.
 
As a first benchmark application to a realistic physical problem,
we have considered the density excitation response in bulk atomic $^{4}$He at $T=0$. The system is modeled 
as a box with periodic boundary conditions, containing $N=64$ or $N=125$ $^{4}$He atoms interacting via the HFDHE2 pair-wise
potential (\cite{Aziz},\cite{Kalos}), which quantitatively reproduces the binding energy of bulk $^{4}$He up 
to the freezing point, effectively including three-body contributions. 
Calculations are performed at the experimental saturation density ($n_{0}=0.02186\;$\AA$^{-3}$). 
The density excitation operator is defined as:
\begin{equation}
\hat{O}(\mathbf{q}) \equiv \hat{\rho} =\sum_{i=1}^{N}e^{i{\mathbf q}\cdot {\mathbf r}_{i}}\,,
\end{equation}
and the transformed DSF in Eq.~($\ref{eq:tr}$) becomes
\[
\Phi(\sigma) =  \langle \Psi_0 \lvert \sum_{i,j=1}^{N}e^{i{\mathbf q}\cdot{\mathbf r}_{i}} K_P(\sigma,\hat H)
e^{-i{\mathbf q}\cdot {\mathbf r}_{j}} \rvert \Psi_0 \rangle\]
As it is customary in neutron spectroscopy one can distinguish the contribution coming 
from the so-called 
{\it coherent} part, given by the terms  with $i\ne j$, related to collective excitations, 
and an {\it incoherent} part with $i=j$ that essentially picks up contributions from single particle excitations.
We   have   obtained results for both the full  and for its incoherent part, 
in the most studied region of the spectrum: the low momentum phonon-maxon-roton part 
$q \approx 0.3 \div 3.5\,$\AA$^{-1}$. 
Computations have been performed by means of a Reptation Monte  Carlo (RMC) algorithm, 
as described in Ref.~\cite{Moroni98}. 
The variational importance function includes two- and three-body correlations expanded in a basis set 
~\cite{Euler} and optimized using a variational Monte Carlo procedure.
Ground state properties are well reproduced: the  ground-state energy per particle is 
$\epsilon_0^{RMC} = -7.23 \pm 0.01$~K, in good agreement with previous calculations 
using the same potential~\cite{Moroni98}, and with the experimental value 
$\epsilon_0^{exp} = -7.17$~K.
The static structure factor $S({\mathbf q})$ is  consistent with 
experimental data and previous calculations (\cite{Moroni98},\cite{Euler}).
\begin{figure}
\centering
\begin{center}
\includegraphics[angle=270,scale=0.35]{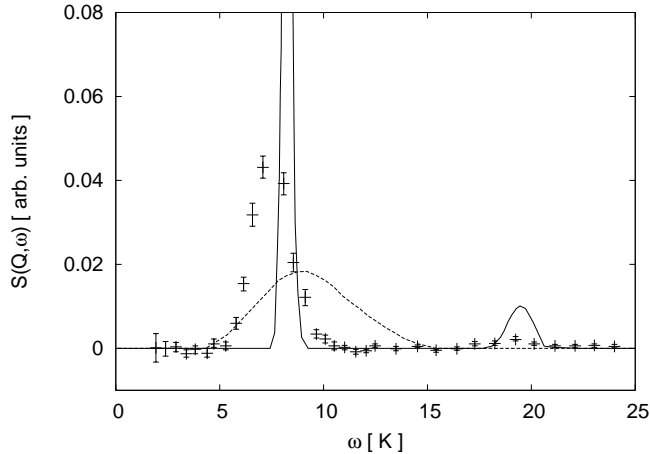}
\end{center}
\caption{A typical result for the response function $(Q,\omega)$ in liquid $^4$He. Points with errorbars are experimental results at $Q=0.4$\AA$^{-1}$ at $T=1.34$K\cite{Glyde}. The continuous line is the result obtained using a Laplace kernel. The dotted line is the result computed by using the Sumudu kernel. Theoretical calculations refer to $Q=0.44$\AA$^{-1}$, value determined by the size of the simulation box.}
\label{fig:low1}
\end{figure}

Turning to the result on $S({\mathbf q},\omega)$, the striking difference between the
estimate obtained inverting the transform with the Laplace kernel or the one in Eq.~(\ref{eq:NEW4})
can be seen in Fig.~\ref{fig:low1}, 
where we compare the results of the inversion obtained from RMC data with both kernels. Apart from 
the small shift of the peak due to the $0.04\,$\AA$^{-1}$ difference in the momentum transfer 
(the momenta  are limited by the discretization imposed
by the use of a finite simulation cell, here L=$14.306 $\AA$ $ ), 
the new kernel permits to retrieve 
the information on the second peak and gives a much more realistic height and width of the one-phonon peak. 

A few comments are necessary concerning the   methods used to invert the transform. 
We have used both the Entropy Maximization Maximum Likelihood (EMML)~\cite{Byrne} and the 
Simultaneous Algebraic Reconstruction Technique (SMART)~\cite{Byrne}. The bars in the figures indicate
the computed widths of the excitations. Both the peak position and the linewidth are robust with respect to the inversion method used. We found that for 
$q \leq 2.4 $\AA$^{-1}$ 
both methods converge to the same solution. 
\begin{figure}
\centering
\includegraphics[angle=270,scale=0.33]{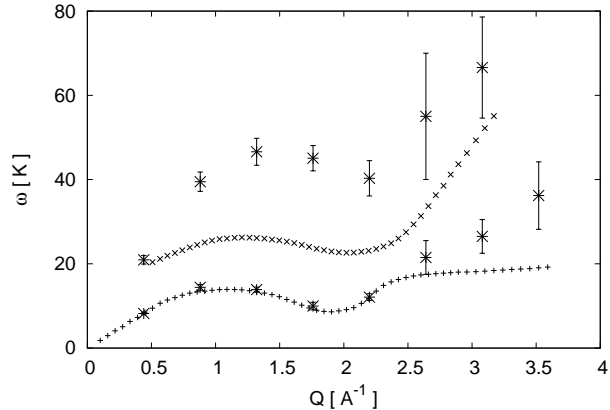}
\caption{Dispersion of the collective modes in liquid $^4$He at equilibrium density and $T=0$. Points with errorbars are the computed values. Errorbars are estimates of the width of the peaks. $+$ and $\times$ are the corresponding experimental data from Ref. \cite{Woods} at $T=1.1$K}
\label{fig:low2}
\end{figure}
In Fig.~\ref{fig:low2} we have plotted the excitation spectrum obtained using the new transform. 
The experimental low-lying part \cite{Glyde} is extremely well reproduced up to $q \approx 2.6$\AA$^{-1}$, where 
the dispersion does not bend over around $2\Delta$ but continues to raise. In this region, however, the 
EMML and SMART give different results indicating that the statistical uncertainty in the QMC data
is too large to allow a consistent reconstruction.
The two-phonon branch is clearly visible and well resolved. As it happens in Ref. \cite{Vitali10}, it
only qualitatively compares to the experiment. 
However, it should be noted that at $T=0$ the peak corresponding to the collective excitation should be 
substantially narrower. An estimate of the intrinsic peak widht is $\Delta\omega\simeq 5\times 10^{-4}$K\cite{Mezei}. Therefore, the experimental width is essentially due to the resolution of the apparatus\cite{Glyde}. 

The dispersion of the peaks of the collective modes, displayed in Fig.2, was obtained combining two simulations at $\Delta \tau = 0.002 K^{-1}$ 
and $0.001 K^{-1}$ respectively obtaining a mesh separation less than $0.5$ untill about $40 K$. In order to obtain meaningful results in the high energy regime, a large collection of RQMC data taken with different imaginary-time steps is needed in order to increase the sampling points. 
\begin{figure}
\centering
\includegraphics[angle=270,scale=0.33]{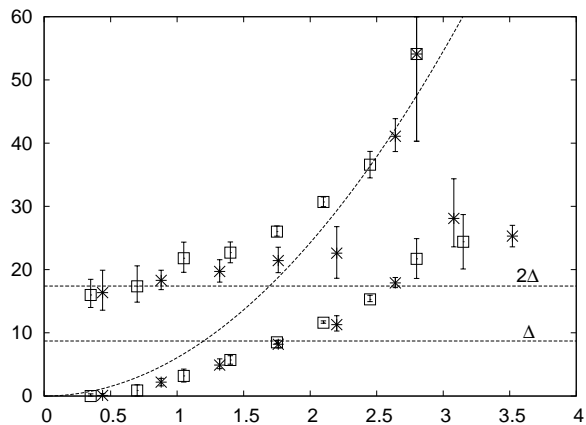}
\caption{Dispersion of the first and second peak of the incoherent DSF computed by means of the Sumudu integral transform. Empty squares are simulations performed in a simulation box with N=64 atoms, stars refer to a box with N=125 atoms. The dashed line is the free-particle excitation spectrum. $\Delta$ is the roton gap, and the lines at energy $\Delta$ and $2\Delta$ are drawn as reference.}
\label{fig:low3}
\end{figure}
Due to this technical difficulty, at present we have not performed an exhaustive
research in the  high momentum-transfer limit. However, some preliminary calculations show that the spectrum has the expected approximately free-particle like behavior, and that for $Q \approx 6 A^{-1}$ and above the incoherent part of the 
Dynamic Structure Factor accounts for the total scattering.

Indeed, the most useful feature of these calculations is that the resolution is 
good enough to allow for separatly computing the  incoherent part of the full response function, in order to study single-particle excitations. 

In Fig.3 we have plotted the calculated excitation spectrum of single-particle excited states. 
The spectrum shows at least two distinct branches. A lower energy excitation starts from $Q \approx 0.5 A^{-1}$ and propagates with a velocity resulting
slightly higher than the superfluid critical velocity $ve/vc \approx 1.57$. A second branch can be observed starting at an energy slightly below two times the roton energy, tending asymptotically to the free particle spectrum.
Interestingly enough, the lower energy branch crosses the collective excitation spectrum exactly at the roton minimum, thereby reinforcing the picture of the roton as a single particle excitation of an atom exiting the superfluid.  The behavior of these single-particle excitations might be significantly affected by the quantum many--body correlations induced by the particle-particle interaction.
An extensive experimental analysis of the single particle excitations properties in superfluid $^4$He is unfortunately not available.

We have proposed a method to extract reliably well resolved spectra from numerical calculations implementing imaginay time propagation of an initial state, such as DMC or RMC. Computations might be easily extended to the $T\neq 0$ case by using standard PIMC methods. The application to the study of the collective and single-particle excitations in $^{4}$He shows the robustness and the higher resolution power of this technique. The limit to the accuracy of the spectra is in principle limited only by the available computer power available.

We thank W. Leidemann, V. Efros, G.V. Chester, M.H. Kalos, J.Carlson and S. Gandolfi for useful and stimulating discussions about this subject. Calculations have been partially carried out on the supercomputer AURORA in the framework of the AURORA-Science project funded by INFN and the Bruno Kessler Foundation (FBK). Authors are members of LISC, Interdisciplinary Laboratory for Computational Science, a joint venture of FBK and the University of Trento.



\end{document}